# Fast and accurate simulations of transmission-line metamaterials using transmission-matrix method

Hui Feng Ma, Tie Jun Cui*, Jessie Yao Chin and Qiang Cheng

Address: The State Key Laboratory of Millimeter Waves, Department of Radio Engineering Southeast University, Nanjing 210096, PR China

Email: Hui Feng Ma - hfma@emfield.org; Tie Jun Cui* - tjcui@seu.edu.cn; Jessie Yao Chin - yaoqin@emfield.org; Qiang Cheng - qiangcheng@emfield.org

* Corresponding author

## Abstract

Recently, two-dimensional (2D) periodically L and C loaded transmission-line (TL) networks have been applied to represent metamaterials. The commercial Agilent's Advanced Design System (ADS) is a commonly-used tool to simulate the TL metamaterials. However, it takes a lot of time to set up the TL network and perform numerical simulations using ADS, making the metamaterial analysis inefficient, especially for large-scale TL networks. In this paper, we propose transmission-matrix method (TMM) to simulate and analyze the TL-network metamaterials efficiently. Compared to the ADS commercial software, TMM provides nearly the same simulation results for the same networks. However, the model-process and simulation time has been greatly reduced. The proposed TMM can serve as an efficient tool to study the TL-network metamaterials.

## Background

In 1968, Veselago first proposed the concept of left-handed material (LHM) [1]. The new medium exhibits many particular properties because of its negative permittivity and negative permeability. However, LHM does not exist in nature, which has to be realized using periodic structures with the unit cell of split-ring resonators (SRR) and conducting wires [2-5] or electric LC (ELC) resonators [6]. The two-dimensional (2D) L and C loaded transmission-line (TL) networks is another method to realize LHM [7-17]. Both isotropic [7-14] and anisotropic TL metamaterials [15-17] have been studied. The particular 2D L and C loaded TL networks can be equivalent to relevant media. In order to obtain the voltage and current distributions on the periodically TL networks, which are equivalent to the field distributions in metamaterial, the Agilent's Advanced Design System (ADS) has been used [8-17]. If we want to observe the node voltages and currents of the LC-loaded TL networks when using ADS, the observation nodes should be labeled in ADS, which is extremely time and memory consuming for large-scale circuit structures. The numerical simulations were time consuming too for large-scale structures using ADS.

Are there some methods which can be used to set up the model quickly and to simulate the TL strucutures efficiently? The transmission-matrix method (TMM) is a good choice [18]. TMM can construct the LC-loaded networks automatically and quickly. The most important thing is that one does not need to label the observation nodes. Hence TMM can save much time and vigor in the construction of TL metamaterials and is more efficient than ADS. Caloz and Itoh have studied such a method in Ref. [19] to analyze TL metamaterials. However, they only considered the periodic structures composed of lumped inductance and capacitance, the composed right/left-handed circuits, in which the transmission lines are not involved [19]. Also, the voltage and current distributions obtained from this method cannot be equivalent to those from the ADS simulations, because the load impedances of observers are added to observation nodes.

In this paper, we generalize the TMM approach to study the 2D LC-loaded TL periodic structures, which have been widely used to represent metamaterials [7-17]. Both the voltage and current distributions on the LC-loaded TL structures can be computed accurately and efficiently using the proposed TMM, which have excellent agreements to the ADS simulation results. However, the time used in the model setup and numerical simulations have been greatly reduced. The proposed TMM approach can also solve large-scale TL metamaterials, which can hardly be realized using ADS.

### *Theoretical analysis of TMM*

The arbitrary 2D TL network under consideration is divided into $m \times n$ unit cells, as shown in Fig. 1(a). The corresponding general T structure of each unit cell is illustrated in Fig. 1(b), in which $Z_0$, $\beta$ and $l$ are the characteristic impedance, the propagation constant, and the length of TL section, respectively; and $Z_s$ and $Y_p$ are loaded inductance and capacitance. Using TMM, we divide the structure into columns of unit cells, one of which is shown as the gray line in Fig. 1(a). The $m \times 1$ unit cells can be represented as a $2m \times 2m$ matrix formed by $m$ input ports and $m$ output ports with the node voltages and port currents [19]. The equivalent column cell with terminated impedance $R_j$ is shown in Fig. 2. The input and output ports are indexed as 1 to $m$ and $m + 1$ to $2m$, respectively, and the corresponding $2m \times 2m$ matrix is expressed as

$$\begin{bmatrix} [V_{in}] \\ [I_{in}] \end{bmatrix} = [T_j] \cdot \begin{bmatrix} [V_{out}] \\ [I_{out}] \end{bmatrix}, \qquad (1)$$

which can be referred to Eq. (4.18) in Ref. [19].

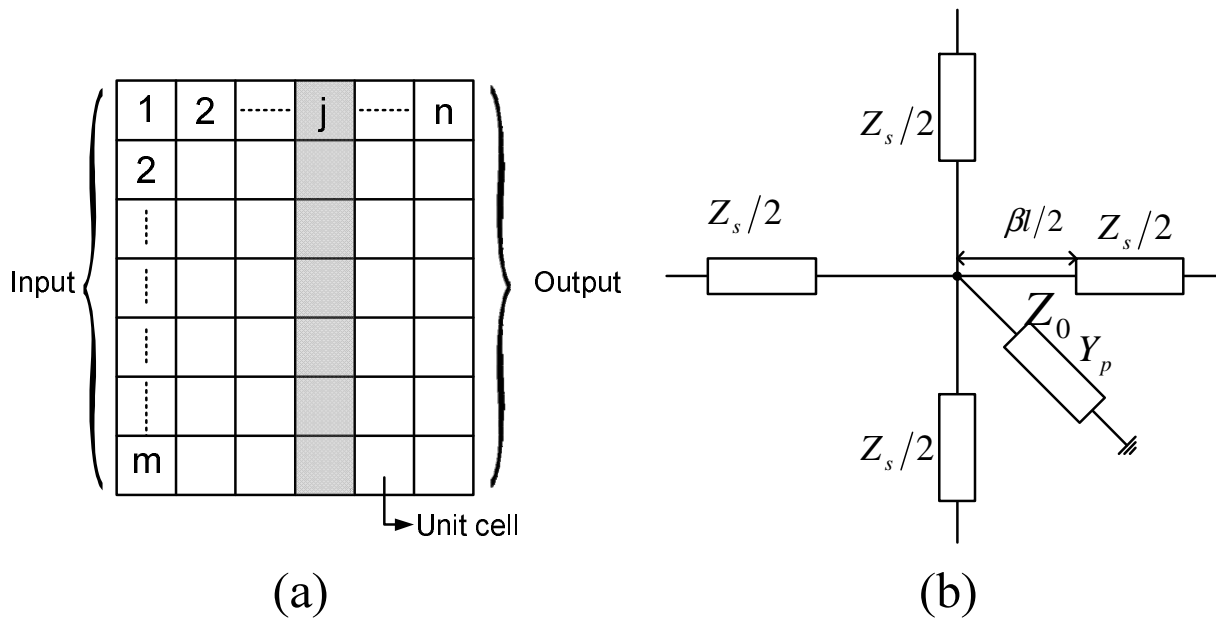

**Figure 1**
(a) Arbitrary 2D L and C loaded TL network. (b) General T network of each unit cell.

By decomposing the column cell into three sub-column cells, as shown in Fig. 2, we can easily obtain the $2m \times 2m$ transmission matrix $[T_j]$ as the product of three matrices $[Th1]$, $[Tv]$ and $[Th2]$:

$$[T_j] = \begin{bmatrix} [A] & [B] \\ [C] & [D] \end{bmatrix} = [Th1] \cdot [T_v] \cdot [Th2], \qquad (2)$$

where the computation of matrices $[Th1]$, $[Tv]$ and $[Th2]$ is similar to that in Ref. [19].

Once $[T_j]$ has been obtained, the *n*-column transmission matrix as shown in Fig. 1(a) will be determined as

$$[T] = \prod_{j=1}^{n} [T_j]. \qquad (3)$$

### *Voltage and current distributions*

We consider the case of an $m \times n$ network with a given source which is arbitrarily loacted at the *k*th row and between the *i*th and $(i + 1)$th columns of the network. We determine to obtain the output voltage and current at any point of the network. The whole network is demonstrated in Fig. 3, where $R'_1, R'_2, \cdots,$ and $R'_m$ are horizontally terminated impedances.

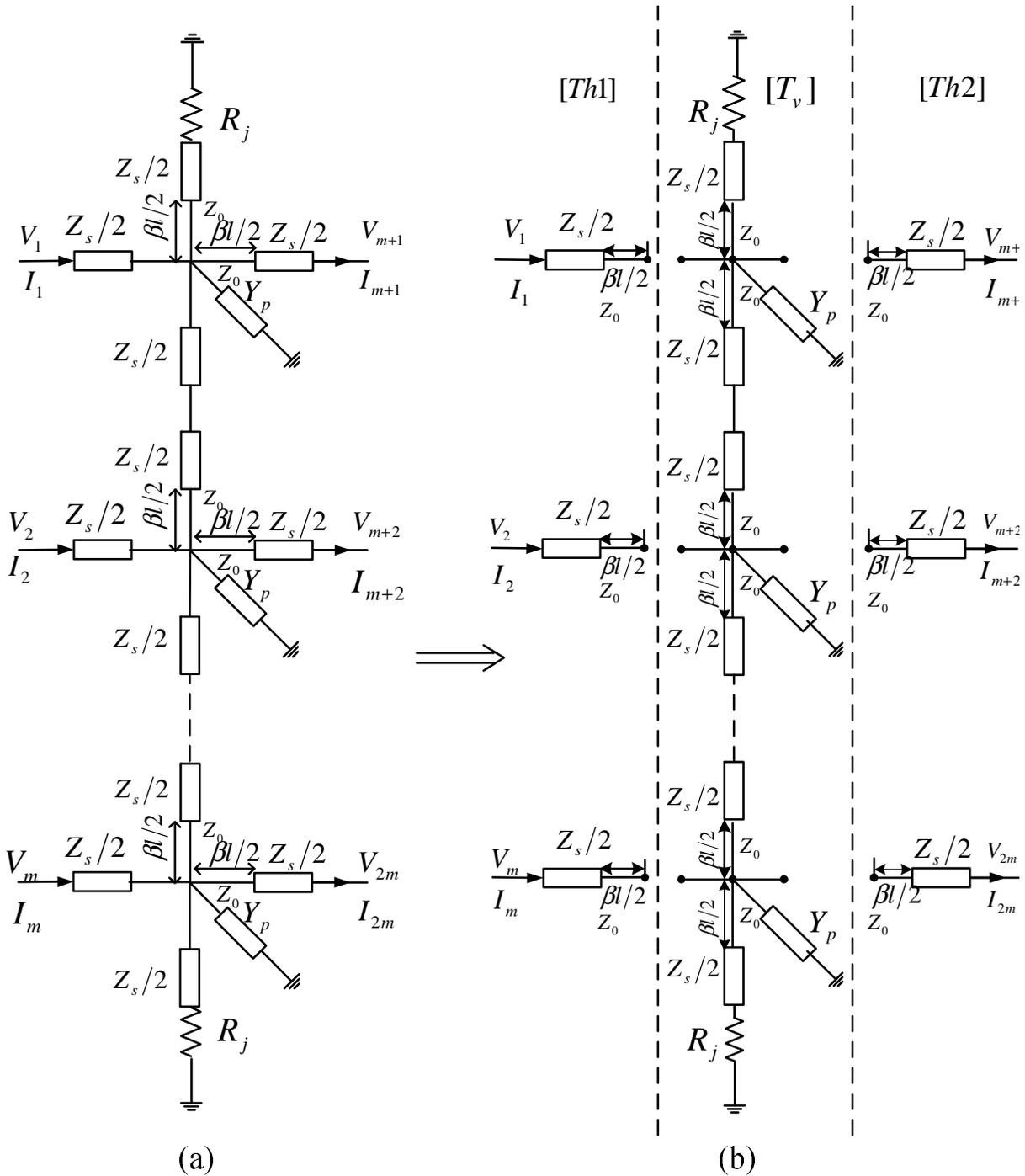

**Figure 2**
The equivalent-circuit T network for a column cell and the corresponding decomposition. (a) T network for a column of cell. (b) The decomposition of three parts.

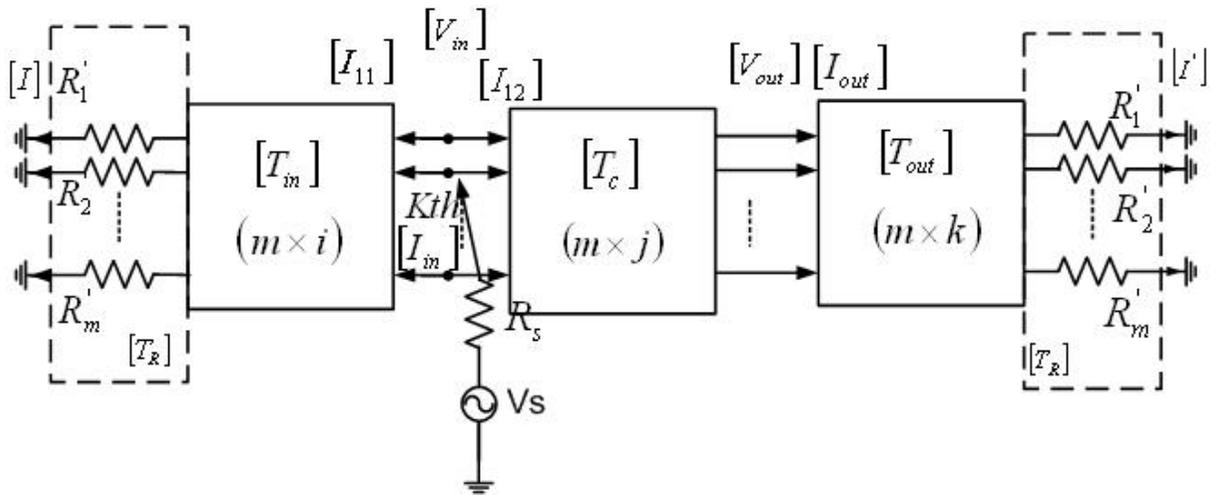

**Figure 3**
The whole network under consideration, including *m* inputs and *m* outputs, which is divided into the input network $T_{in}$, the center network $T_c$, and the output network $T_{out}$. A source is arbitrarily located at the *k*th row and between the *i*th and (*i* + 1)th columns of the network.

Let

$$[T'_{in}] = [T_{in}] \cdot [T_R] = \begin{bmatrix} [A'_{in}] & [B'_{in}] \\ [C'_{in}] & [D'_{in}] \end{bmatrix}, \quad (4)$$

$$[T'_{out}] = [T_c] \cdot [T_{out}] \cdot [T_R] = \begin{bmatrix} [A'_{out}] & [B'_{out}] \\ [C'_{out}] & [D'_{out}] \end{bmatrix}, \quad (5)$$

in which

$$[T_R] = \begin{bmatrix} [I] & [X] \\ [O] & [I] \end{bmatrix}, \quad (6)$$

and

$$[X] = \begin{bmatrix} R'_1 & 0 & \cdots & 0 \\ 0 & R'_2 & \cdots & 0 \\ \vdots & \vdots & \ddots & \vdots \\ 0 & 0 & 0 & R_m \end{bmatrix}. \quad (7)$$

In Eqs. (4) and (5), $[A'_{in}], [B'_{in}], [C'_{in}], [D'_{in}], [A'_{out}], [B'_{out}], [C'_{out}], [D'_{out}]$ and $[X]$ are all $m \times m$ matrices.

Using the series transmission-matrix theory, we can easily obtain the following equations

$$\begin{bmatrix} [V_{in}] \\ [I_{11}] \end{bmatrix} = [T'_{in}] \cdot \begin{bmatrix} [O] \\ [I] \end{bmatrix}, \quad (8)$$

$$\begin{bmatrix} [V_{in}] \\ [I_{12}] \end{bmatrix} = [T'_{out}] \cdot \begin{bmatrix} [O] \\ [I'] \end{bmatrix}, \tag{9}$$

here, $[O]$ is $m \times 1$ zero matrix, $[I]$ and $[I']$ are the current as illustrates in Fig. 3. From Eqs. (8) and (9), we have

$$[I_{11}] = [D'_{in}] \cdot [B'_{in}]^{-1} \cdot [V_{in}] = [Y'_{in}] \cdot [V_{in}], \tag{10}$$

$$[I_{12}] = [D'_{out}] \cdot [B'_{out}]^{-1} \cdot [V_{in}] = [Y'_{out}] \cdot [V_{in}], \tag{11}$$

and

$$[I_{in}] = [I_{11}] + [I_{12}]. \tag{12}$$

Furthermore, we obtain

$$[V_{in}] = ([Y'_{in}] + [Y'_{out}])^{-1} \cdot [I_{in}] = [Y] \cdot [I_{in}]. \tag{13}$$

According to Eq. (4.32) in Ref. [19], the input current $[I_{in}]_{m \times 1}$ will be zero except the $k$th-row current $I_{in, k}$. The $k$th-row input voltage $V_{in, k} = V_s - R_s \cdot I_{in, k}$. Then we have

$$I_{in,k} = \frac{V_s}{R_s + \gamma_{kk}}, \tag{14}$$

in which $\gamma_{kk}$ is the $k$th-row and $k$th-column element of matrix $[Y]$.

We remark that the $k$th element of the matrix $[I_{in}]$ is $I_{in, k}$, and all other elements are zero. Hence, once we obtain $I_{in, k}$, the matrix $[I_{in}]$ will be known, and then matrices $[V_{in}]$, $[I_{11}]$ and $[I_{12}]$ will also be determined from Eqs. (13), (10) and (11), respectively.

Once we know $[V_{in}]$, $[I_{11}]$ and $[I_{12}]$, the output voltage and current at any point of the network which is located on the right side of the source will be determined easily from the following equation

$$\begin{bmatrix} [V_{out}] \\ [I_{out}] \end{bmatrix} = [T_C]^{-1} \cdot \begin{bmatrix} [V_{in}] \\ [I_{12}] \end{bmatrix}. \tag{15}$$

Similarly, we can obtain the output voltages and currents at points located on the left side of the source. Using the same method, we can also deal with the cases of two given sources which are located at the same row.

The above method can be used to analyze the electromagnetic properties of 2D TL metamaterials such as super lens [7-11], EM localizations [12,13], and super waveguides [14] with high efficiencies and high accuracies.

*Simulation results*

In order to verify the correctness of the proposed TMM method, we re-compute the sub-wavelength focusing structure possessing the same TL network and parameters as those in Ref. [9] using TMM. The corresponding TMM-simulation results are shown in Figs. 4 and 5. Comparing with the ADS simulations [9], we notice that two results have excellent agreements in both amplitude and phase distributions. The results show that it is valid to replace ADS by TMM in numerical simulations of TL metamaterial structures. In this example, TMM does not exhibit its advantage obviously in saving time with the small-scale network.

To verify the accuracy of TMM further, we give a full comparison of TMM-simulation results to the commercial ADS-simulation results. We consider an example of super waveguide, which is a planar waveguide filled with air and left-handed material (LHM) [14,20]. It has been shown that extremely high-power densities with opposite propagation directions can be generated and transmitted along the waveguide if the air and LHM have equal thickness and the permittivity and permeability of LHM satisfy $\varepsilon = \varepsilon_0(1 + \delta)$ and $\mu = -\mu_0/(1 + \delta)$, in which $\delta$ is a small parameter [20]. The above theoretical prediction has been realized using 2D LC-loaded TL metamaterials [14]. Now we re-analysis the TL-metamaterial super waveguide using the TMM technique.

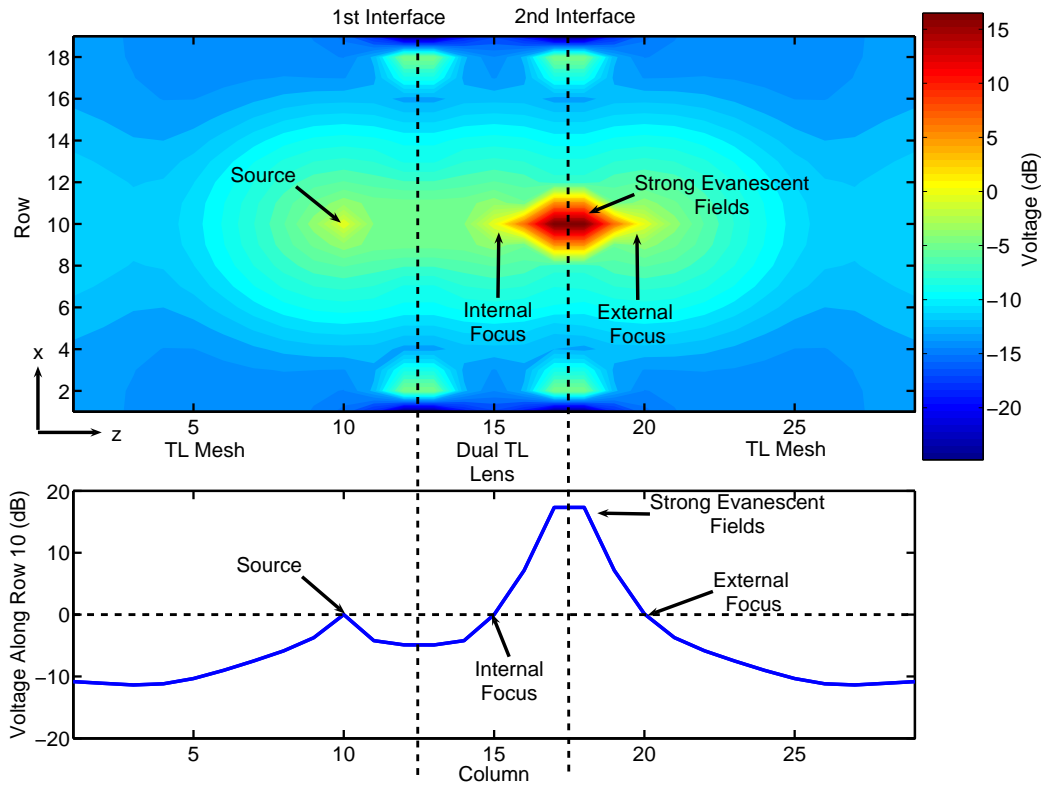

**Figure 4**
TMM-simulation results for TL super lens [9] at 1 GHz (voltage amplitude.

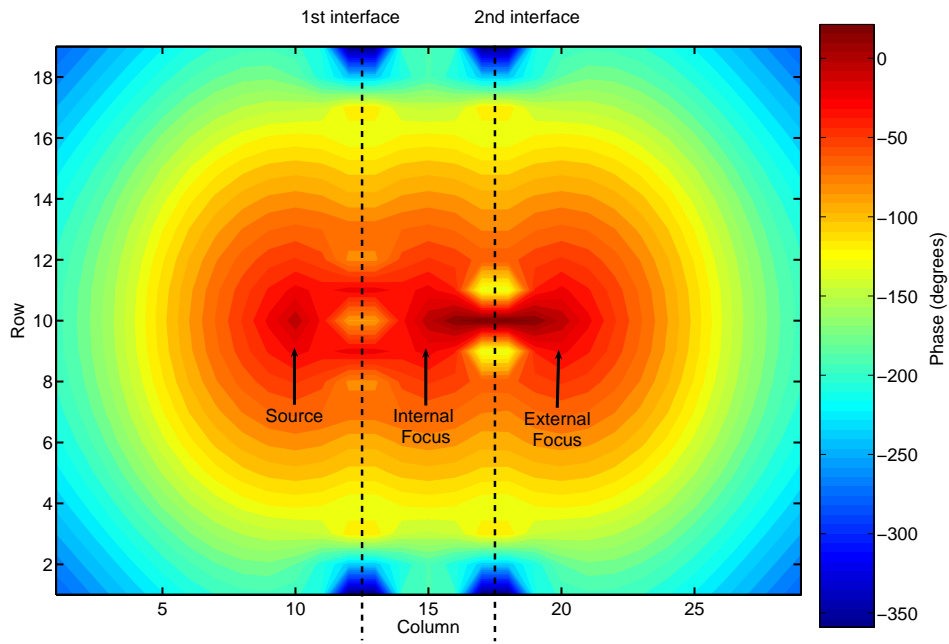

**Figure 5**
TMM-simulation results for TL super lens [9] at 1 GHz (voltage phase).

Consider a TL-metamaterial super waveguide which is composed of 179 × 18 unit cells [14], where the right-handed TL (RHTL) and the left-handed TL (LHTL) parts extend 9 cells in the *x* direction and 179 cells in the *y* direction, respectively. A 1-V (0 dB) voltage source is connected to the node of the cell numbered as (4, 90) in the RHTL region. We choose the size of unit cell as $d$ = 1 cm, the small parameter as $\delta$ = -0.053, the characteristic impedance in TL as $Z_0$ = 533.1459 Ω, and the propagation constant in TL as $\beta$ = 14.8096. The simulation results of voltage and current distributions computed by TMM are illustrated in Fig. 6. As a comparison, the corresponding voltage distributions simulated by ADS are shown in Fig. 7. From these two figures, excellent agreements have been observed for the voltage distributions. The current distributions have also excellent agreements, which are not repeated here.

We remark that much less time has been used in TMM than that in ADS, since it takes long time to label the voltage and current probes in ADS for the 179 × 18 network. At least several days were taken to set up and simulate such a large-scale TL metamaterial using ADS, while it only takes several minutes to solve the same problem using TMM.

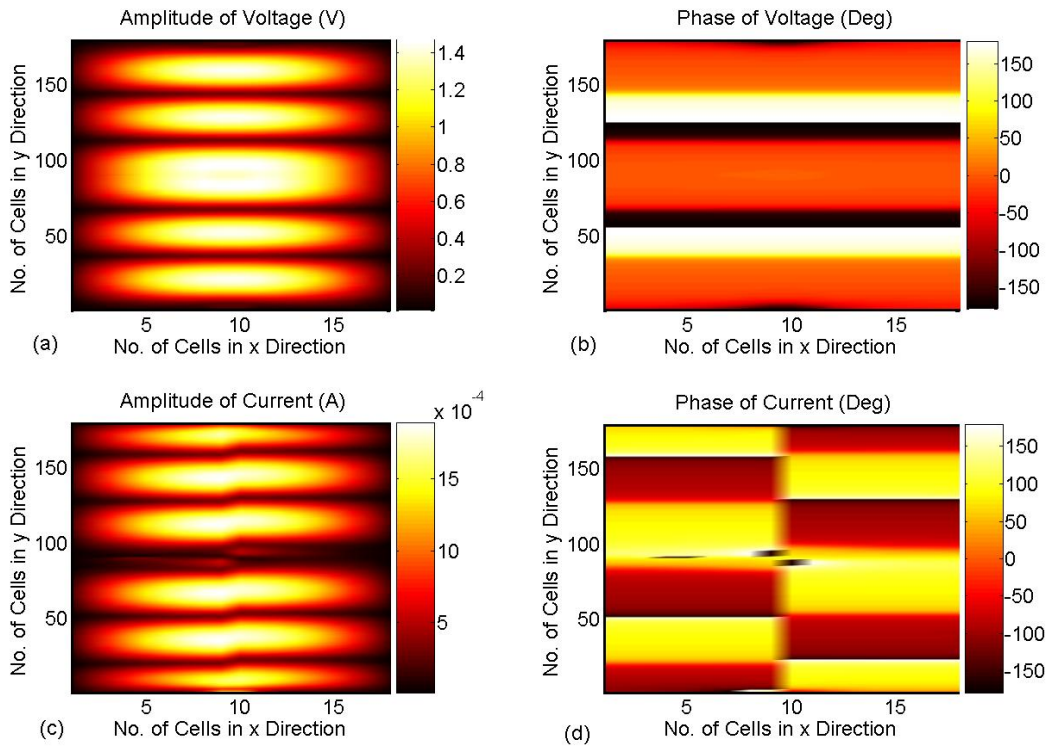

**Figure 6**
The voltage and current distributions on the planar TL-metamaterial super waveguide with 18 × 179 unit cells computed by TMM. (a) The amplitude distribution of voltage. (b) The phase distribution of voltage. (c) The amplitude distribution of current. (d) The phase distribution of current.

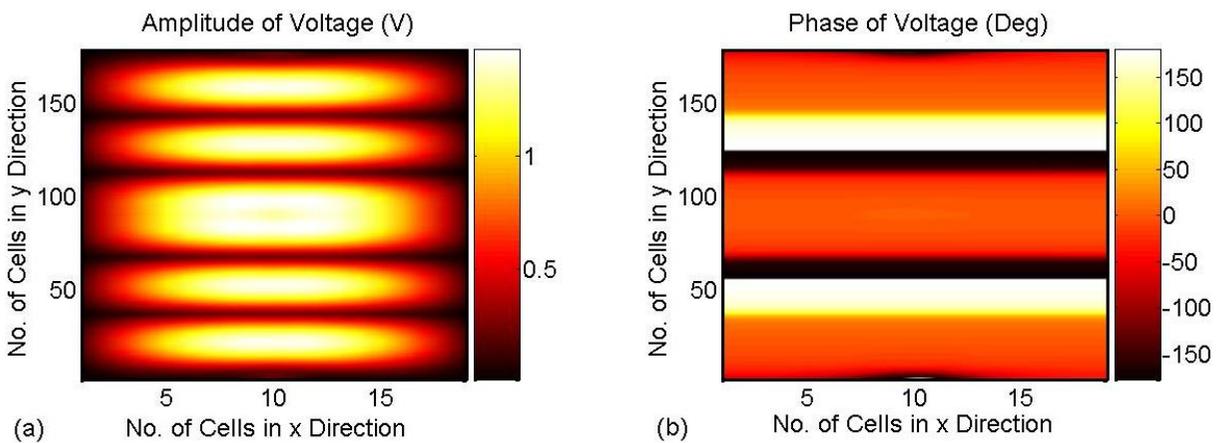

**Figure 7**
The voltage and current distributions on the planar TL-metamaterial super waveguide with 18 × 179 unit cells simulated by the ADS commercial software. (a) The amplitude distribution of voltage. (b) The phase distribution of voltage.

To further verify the accuracy of TMM, the voltage and current distributions on the planar TL-metamaterial super waveguide along the *x* direction with *y* = 90 are computed using TMM and ADS, as shown in Fig. 8. Clearly, the two results have excellent agreements.

Then we consider a longer TL-metamaterial super waveguide, which is composed of 599 × 18 unit cells. Similar to the previous case, the RHTL and LHTL parts extend 9 cells in the *x* direction but 599 cells in the *y* direction. A 1-V (0-dB) voltage source is located at the node numbered as (4, 300). The TMM-simulation results for the voltage and current distributions are shown in Fig. 9. We remark that ADS cannot be used to handle the problem in a 2GB-memory personal computer due to the large requirement of memory.

In another example, we consider an LHTL superlens and use it to localize electromagnetic waves and energies [12,13]. To build up the TL-metamaterial superlens [12], 33 × 21 unit cells have been used, where the left RHTL part, the middle LHTL part, and the right RHTL part extend to 7 cells in the *x* direction and 33 unit cells in the *y* direction, respectively. Two 1-V (0 dB) voltage sources $V_1$ and $V_2$ with opposite phases are connected to the nodes numbered as (17, 4) and (17,

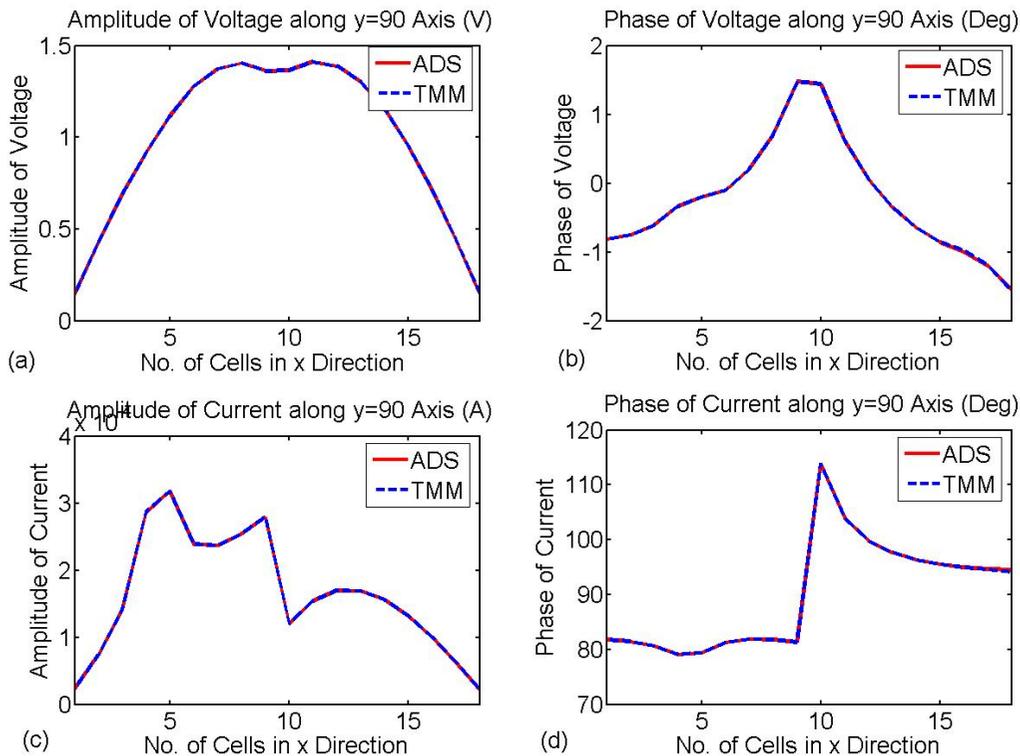

#### Figure 8
The comparison of voltage and current distributions on the planar TL-metamaterial super waveguide along the line *y* = 90, where red lines indicate the ADS simulation results, and the dashed blue lines are TMM computation results. (a) The amplitude of voltage. (b) The phase of voltage. (c) The amplitude of current. (d) The phase of current.

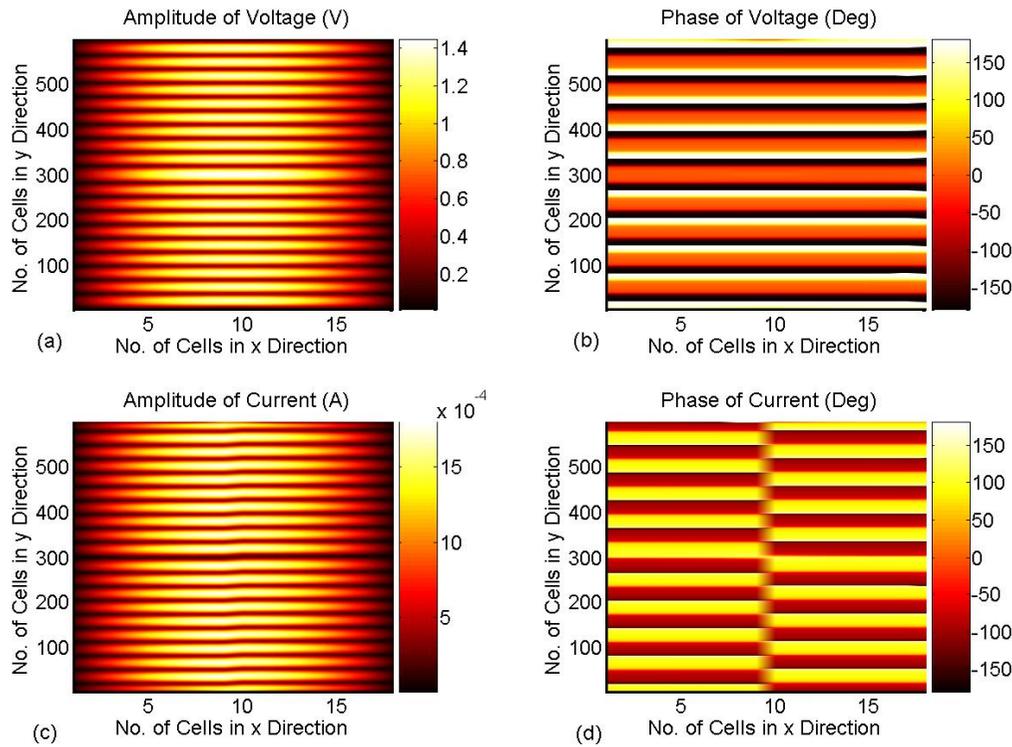

**Figure 9**
The voltage and current distributions of a longer TL-metamaterial super waveguide with 18 × 599 unit cells computed by TMM. For the same structure, ADS cannot be used due to the memory limit of personal computer. (a) The amplitude distribution of voltage. (b) The phase distribution of voltage. (c) The amplitude distribution of current. (d) The phase distribution of current.

17) in the RHTL regions, respectively. Here, we chose $d$ = 1 mm, $\delta$ = $10^{-4}$, the characteristic impedance $Z_0$ = 533.1459 Ω, the propagation constant $\beta$ = 14.8096, the series capacitance $C$ = 10.0773 pF, and the shunt inductance $L$ = 1432.2120 nH. The TMM-simulation results of the voltage and current distributions are demonstrated in Fig. 10, and the corresponding ADS-simulation results are shown in Fig. 11. Comparing Figs. 10 and 11, such two results have excellent agreements.

Similar to the earlier example, we have observed the voltage and current distributions computed by the two methods along a line in the $x$ direction with $y$ = 17, as illustrated in Fig. 12. The results demonstrate that TMM and ADS provide nearly the same accuracy.

We also consider a large-scale TL-metamaterial superlens for two sources with 599 × 21 unit cells, which cannot be solved using ADS. In this example, two 1-V (0 dB) voltage sources with opposite phases are connect to the nodes numbered as (4, 300) and (17, 300) in the RHTL regions, respectively. The TMM-simulation results of the voltage and current distributions are illustrated in Fig. 13.

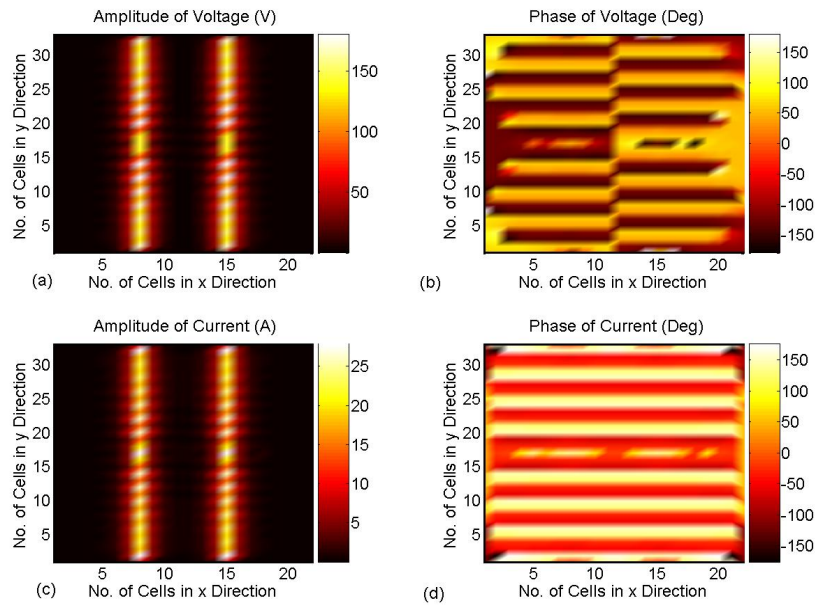

**Figure 10**
The voltage and current distributions on the TL-metamaterial superlens for two sources computed by TMM, where 21 × 33 unit cells have been used. (a) The amplitude distribution of voltage. (b) The phase distribution of voltage. (c) The amplitude distribution of current. (d) The phase distribution of current.

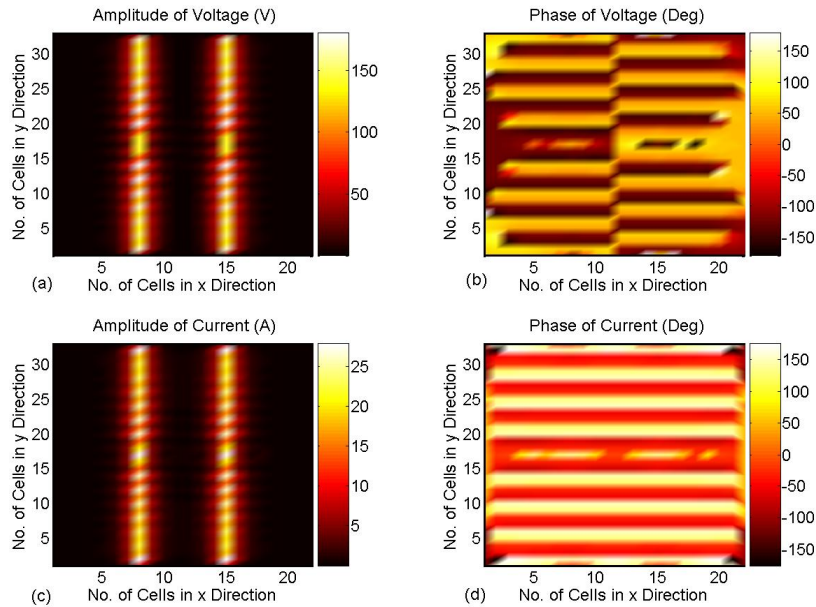

**Figure 11**
The voltage and current distributions on the TL-metamaterial superlens for two sources simulated by ADS, where 21 × 33 unit cells have been used. (a) The amplitude distribution of voltage. (b) The phase distribution of voltage. (c) The amplitude distribution of current. (d) The phase distribution of current.

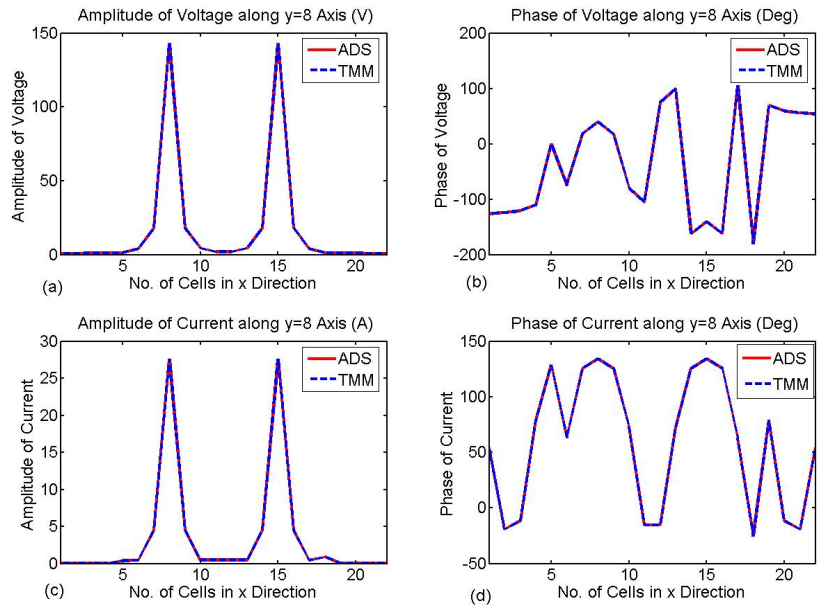

**Figure 12**
The amplitudes and phases of voltages and currents along the line $y = 17$ in the superlens for two sources, where the red lines are ADS simulation results and the dashed blue lines are TMM computation results. (a) The amplitudes of voltages. (b) The phases of voltages. (c) The amplitudes of currents. (d) The phases of currents.

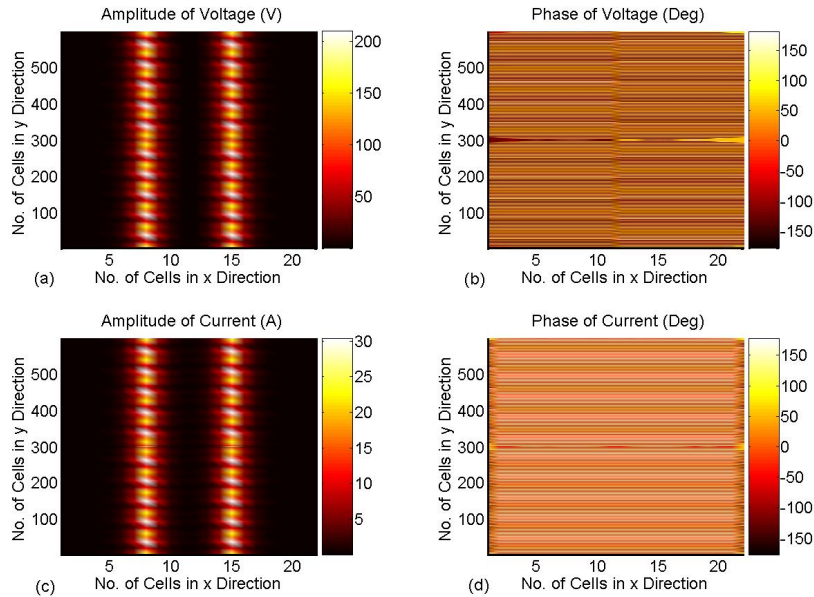

**Figure 13**
The voltage and current distributions on the long TL-metamaterial superlens for two sources computed by TMM, where $21 \times 599$ unit cells have been used. (a) The amplitude distribution of voltage. (b) The phase distribution of voltage. (c) The amplitude distribution of current. (d) The phase distribution of current.

**Conclusion**

In this paper, a TMM approach has been proposed to determine the voltage and current distributions on 2D LC-loaded TL metamaterials. The proposed TMM has the same accuracy as ADS for numerical simulations of finite-size metamaterial structures. However, it is much easier to set up LC-loaded TL models and save much more time using TMM than ADS. For large-scale metamaterial structures, ADS can even not be used due to the large memory requirements. As an example, for the large-scale TL-metamaterial super waveguide [14] with 179 × 18 unit cells, whose results are shown in Figs. 6 and 7, the total computation time is only 36.543 seconds using TMM in a 2G-memory personal computer. The method is realized by MATLAB with computing 18 358 × 358 matrices. We also realized the same super waveguide using ADS. It took us nearly two weeks to set up the circuit model and label the observation nodes in the same 2G-memory personal computer. The boring work is labeling the observation nodes, which must be very careful and to avoid mistakes. When the circuit structure is constructed, the total simulation time using ADS is about fifteen minutes. Compared to ADS, TMM is much more efficient when treating particular circuits with large amount of same unit cells such as the TL metamaterials.

**Acknowledgements**

This work was supported in part by the National Basic Research Program (973) of China under Grant No. 2004CB719802, in part by the National Science Foundation of China under Grant Nos. 60671015, 60496317, 60225001, and 60621002, and in part by the National Doctoral Foundation of China under Grant No. 20040286010.